\documentclass[10pt,draft]{article}
\usepackage{amsfonts,amsmath,amssymb,cite}
\newcommand{\Real}{\mathop{\textrm{Re}}}
\newcommand{\sgn}{\mathop{\textrm{sgn}}}

\begin{document}
\title{Electric and magnetic dipole shielding constants for the
ground state of the relativistic hydrogen-like atom: Application of
the Sturmian expansion of the generalized Dirac--Coulomb Green 
function}
\author{Patrycja Stefa{\'n}ska 
and 
Rados{\l}aw Szmytkowski\footnote{Corresponding author. 
Email: radek@mif.pg.gda.pl} \\*[3ex]
Atomic Physics Division,
Department of Atomic Physics and Luminescence, \\
Faculty of Applied Physics and Mathematics,
Gda{\'n}sk University of Technology, \\
Narutowicza 11/12, 80--233 Gda{\'n}sk, Poland}
\date{\today}
\maketitle
\begin{abstract}
The Sturmian expansion of the generalized Dirac--Coulomb Green
function [R.~Szmytkowski, J.\ Phys.\ B 30 (1997) 825; erratum 30
(1997) 2747] is exploited to derive closed-form expressions for
electric ($\sigma_{\mathrm{E}}$) and magnetic ($\sigma_{\mathrm{M}}$)
dipole shielding constants for the ground state of the relativistic
hydrogen-like atom with a point-like and spinless nucleus of charge
$Ze$. It is found that $\sigma_{\mathrm{E}}=Z^{-1}$ (as it should be)
and $$\sigma_{\mathrm{M}}=-(2Z\alpha^{2}/27)
(4\gamma_{1}^{3}+6\gamma_{1}^{2}-7\gamma_{1}-12)
/[\gamma_{1}(\gamma_{1}+1)(2\gamma_{1}-1)],$$ where
$\gamma_{1}=\sqrt{1-(Z\alpha)^{2}}$ ($\alpha$ is the fine-structure
constant). This expression for $\sigma_{\mathrm{M}}$ agrees with
earlier findings of several other authors, obtained with the use of
other analytical techniques, and is elementary compared to an
alternative one presented recently by Cheng \emph{et al.\/} [J.\
Chem.\ Phys.\ 130 (2009) 144102], which involves an infinite series
of ratios of the Euler's gamma functions.
\end{abstract}
%
%\newpage
%
\section{Introduction}
\label{I}
\setcounter{equation}{0}
In the series of papers published by our group over the period of
past several years, it has been shown that the Sturmian expansion of
the generalized (or reduced) Dirac--Coulomb Green function (GDCGF),
found in Ref.\ \cite{Szmy97}, may be used as a convenient tool in
perturbation-theory calculations of some electromagnetic properties
of relativistic one-electron atoms. In particular, closed-form
expressions, in terms of the generalized hypergeometric function
$_{3}F_{2}$ with the unit argument, have been derived for the
ground-state static dipole magnetizability \cite{Szmy02}, the
polarizability \cite{Szmy04} and the Stark-induced magnetic anapole
moment \cite{Miel06} for the system.

Recently, Cheng \emph{et al.\/} \cite{Chen09} have reported the use
of the GDCGF Sturmian expansion technique of Ref.\ \cite{Szmy97} for
the purpose to find the magnetic dipole shielding constant
$\sigma_{\mathrm{M}}$ for the Dirac one-electron atom in its ground
state. Absolutely no details of calculations have been provided in
Ref.\ \cite{Chen09}; only the final expression for
$\sigma_{\mathrm{M}}$ has been given therein as a sum of two
contributions, one being elementary and the second one having a form
of an infinite series of ratios of the Euler's gamma functions. A
literature search shows that calculations of $\sigma_{\mathrm{M}}$
for the same system were carried out before by several research
groups and published in Refs.\ 
\cite{Zapr74,Zapr81,Zapr85,Moor99,Pype99,Ivan09} (none of those works
has been referenced in Ref.\ \cite{Chen09}). An expression for
$\sigma_{\mathrm{M}}$ arrived at in Refs.\
\cite{Moor99,Pype99,Ivan09} (a corresponding formula given in Refs.\
\cite{Zapr74,Zapr81,Zapr85} contains a misprint, cf.\ Sec.\ \ref{III}
below) appears to be elementary compared to the one in Ref.\
\cite{Chen09}. This prompts one to inquire whether the GDCGF Sturmian
expansion technique is practically capable to provide the same simple
representation of $\sigma_{\mathrm{M}}$ as the one given in Refs.\
\cite{Moor99,Pype99,Ivan09} (and, after a due correction is made,
also in Refs.\ \cite{Zapr74,Zapr81,Zapr85}). In the present paper, we
answer this question affirmatively.

When preparing this report, we have decided not to focus only on the
dipole magnetic shielding constant $\sigma_{\mathrm{M}}$ for the
Dirac one-electron atom in the ground state, but to present at first
in Sec.\ \ref{II} details of calculations of the electric dipole
shielding constant $\sigma_{\mathrm{E}}$ for the same system. The
value of the latter quantity is known exactly to be $Z^{-1}$
\cite{Ster54}, where $Z$ is the nuclear charge in the units of $e$.
This fact makes the evaluation of $\sigma_{\mathrm{E}}$ an ideal test
of correctness and robustness of any analytical technique, including
the present one. We believe the material of Sec.\ \ref{II} is highly
instructive, as it shows how certain infinite series encountered in
calculations based on the GDCGF Sturmian expansion technique may be
summed to closed elementary forms. The practical knowledge gained in
that way is then successfully exploited in calculations of
$\sigma_{\mathrm{M}}$ reported in Sec.\ \ref{III}.

A small part of the material of Sec.\ \ref{III} has been presented in
an unpublished comment \cite{Szmy11} on Ref.\ \cite{Chen09}.
%
%\newpage
%
\section{The electric dipole shielding constant}
\label{II}
\setcounter{equation}{0}
Consider a Dirac one-electron atom with an infinitely heavy,
point-like and spinless nucleus of charge $Ze$. In the presence of a
weak, static, uniform electric field $\boldsymbol{E}_{\mathrm{ext}}$,
an electronic wave function of a ground quasi-bound state is, to the
first order in the perturbing field, approximated by
\begin{equation}
\Psi(\boldsymbol{r})\simeq\Psi^{(0)}(\boldsymbol{r})
+\Psi^{(1)}(\boldsymbol{r}).
\label{2.1}
\end{equation}
Here, $\Psi^{(0)}(\boldsymbol{r})$ is the ground-state wave function
of an isolated atom and is given by
\begin{equation}
\Psi^{(0)}(\boldsymbol{r})
=a_{1/2}^{(0)}\Psi_{1/2}^{(0)}(\boldsymbol{r})
+a_{-1/2}^{(0)}\Psi_{-1/2}^{(0)}(\boldsymbol{r}),
\label{2.2}
\end{equation}
where
\begin{equation}
|a_{1/2}^{(0)}|^{2}+|a_{-1/2}^{(0)}|^{2}=1
\label{2.3}
\end{equation}
(otherwise the coefficients $a_{\pm1/2}^{(0)}$ are arbitrary),
\begin{equation}
\Psi_{\mu}^{(0)}(\boldsymbol{r})
=\frac{1}{r}
\left(
\begin{array}{c}
P^{(0)}(r)\Omega_{-1\mu}(\boldsymbol{n}_{r}) \\
\mathrm{i}Q^{(0)}(r)\Omega_{1\mu}(\boldsymbol{n}_{r}) \\
\end{array}
\right)
\qquad ({\textstyle\mu=\pm\frac{1}{2}}),
\label{2.4}
\end{equation}
with
\begin{equation}
P^{(0)}(r)=-\sqrt{\frac{Z}{a_{0}}\frac{1+\gamma_{1}}
{\Gamma(2\gamma_{1}+1)}}
\left(\frac{2Zr}{a_{0}}\right)^{\gamma_{1}}\exp(-Zr/a_{0}),
\label{2.5}
\end{equation}
\begin{equation}
Q^{(0)}(r)=\sqrt{\frac{Z}{a_{0}}\frac{1-\gamma_{1}}
{\Gamma(2\gamma_{1}+1)}}
\left(\frac{2Zr}{a_{0}}\right)^{\gamma_{1}}\exp(-Zr/a_{0}),
\label{2.6}
\end{equation}
and with $\Omega_{\kappa\mu}(\boldsymbol{n}_{r})$,
($\boldsymbol{n}_{r}=\boldsymbol{r}/r$), being the orthonormal
spherical spinors defined as in Ref.\ \cite{Szmy07}. The second term
on the right-hand side of Eq.\ (\ref{2.1}),
$\Psi^{(1)}(\boldsymbol{r})$, is the first-order perturbation-theory
correction to $\Psi^{(0)}(\boldsymbol{r})$ given by
\begin{equation}
\Psi^{(1)}(\boldsymbol{r})
=-e\boldsymbol{E}_{\mathrm{ext}}\cdot\int_{\mathbb{R}^{3}}
\mathrm{d}^{3}\boldsymbol{r}'\:
\bar{G}\mbox{}^{(0)}(\boldsymbol{r},\boldsymbol{r}')
\boldsymbol{r}'\Psi^{(0)}(\boldsymbol{r}'),
\label{2.7}
\end{equation}
with $\bar{G}\mbox{}^{(0)}(\boldsymbol{r},\boldsymbol{r}')$ being the
generalized Dirac--Coulomb Green function associated with the
ground-state hydrogenic energy level 
$\mathcal{E}^{(0)}=mc^{2}\gamma_{1}$. 

The symbol $\gamma_{\kappa}$, appearing in Eqs.\ (\ref{2.5}),
(\ref{2.6}), and also in later considerations, is standardly defined
as
\begin{equation}
\gamma_{\kappa}=\sqrt{\kappa^{2}-(Z\alpha)^{2}},
\label{2.8}
\end{equation}
where $\alpha$ is the Sommerfeld's fine-structure constant. Moreover,
as usual, $a_{0}$ denotes the Bohr radius.

The electric field produced at the point $\boldsymbol{r}$ by the
atomic electron, the latter being in the state characterized by the
wave function $\Psi(\boldsymbol{r})$, is given by
\begin{equation}
\boldsymbol{E}(\boldsymbol{r})=\frac{1}{4\pi\epsilon_{0}}
\int_{\mathbb{R}^{3}}\mathrm{d}^{3}\boldsymbol{r}'\:
\rho_{\mathrm{e}}(\boldsymbol{r}')
\frac{\boldsymbol{r}-\boldsymbol{r}'}
{|\boldsymbol{r}-\boldsymbol{r}'|^{3}},
\label{2.9}
\end{equation}
where
\begin{equation}
\rho_{\mathrm{e}}(\boldsymbol{r})=-e\Psi^{\dag}(\boldsymbol{r})
\Psi(\boldsymbol{r})
\label{2.10}
\end{equation}
is the smeared electronic charge density distribution. Hence, at the
point $\boldsymbol{r}=\boldsymbol{0}$, where the nucleus is located,
the field is
\begin{equation}
\boldsymbol{E}(\boldsymbol{0})=-\frac{1}{4\pi\epsilon_{0}}
\int_{\mathbb{R}^{3}}\mathrm{d}^{3}\boldsymbol{r}'\:\boldsymbol{r}'
\frac{\rho_{\mathrm{e}}(\boldsymbol{r}')}{r^{\prime\,3}}.
\label{2.11}
\end{equation}
Now, if $\Psi(\boldsymbol{r})$ is approximated as in Eq.\
(\ref{2.1}), we have
\begin{equation}
\rho_{\mathrm{e}}(\boldsymbol{r})
\simeq\rho_{\mathrm{e}}^{(0)}(\boldsymbol{r})
+\rho_{\mathrm{e}}^{(1)}(\boldsymbol{r}),
\label{2.12}
\end{equation}
where
\begin{equation}
\rho_{\mathrm{e}}^{(0)}(\boldsymbol{r})
=-e\Psi^{(0)\dag}(\boldsymbol{r})
\Psi^{(0)}(\boldsymbol{r})
\label{2.13}
\end{equation}
and
\begin{equation}
\rho_{\mathrm{e}}^{(1)}(\boldsymbol{r})
=-2e\Real[\Psi^{(0)\dag}(\boldsymbol{r})\Psi^{(1)}(\boldsymbol{r})].
\label{2.14}
\end{equation}
Consequently, $\boldsymbol{E}(\boldsymbol{0})$ is approximately given
by
\begin{equation}
\boldsymbol{E}(\boldsymbol{0})
\simeq\boldsymbol{E}^{(0)}(\boldsymbol{0})
+\boldsymbol{E}^{(1)}(\boldsymbol{0}),
\label{2.15}
\end{equation}
with
\begin{equation}
\boldsymbol{E}^{(0)}(\boldsymbol{0})=\frac{e}{4\pi\epsilon_{0}}
\int_{\mathbb{R}^{3}}\mathrm{d}^{3}\boldsymbol{r}'\:
\boldsymbol{r}'\frac{\Psi^{(0)\dag}(\boldsymbol{r}')
\Psi^{(0)}(\boldsymbol{r}')}{r^{\prime\,3}}
\label{2.16}
\end{equation}
and
\begin{equation}
\boldsymbol{E}^{(1)}(\boldsymbol{0})=\frac{2e}{4\pi\epsilon_{0}}
\Real\int_{\mathbb{R}^{3}}\mathrm{d}^{3}\boldsymbol{r}'\:
\boldsymbol{r}'\frac{\Psi^{(0)\dag}(\boldsymbol{r}')
\Psi^{(1)}(\boldsymbol{r}')}{r^{\prime\,3}}.
\label{2.17}
\end{equation}
It may be easily shown that, because of parity reasons, the field
component at $\boldsymbol{r}=\boldsymbol{0}$ due to the unperturbed
electronic distribution, $\boldsymbol{E}^{(0)}(\boldsymbol{0})$,
vanishes. 

In turn, using the expression (\ref{2.7}) for
$\Psi^{(1)}(\boldsymbol{r})$, we find that the first-order field 
correction $\boldsymbol{E}^{(1)}(\boldsymbol{0})$ may be written in
the form
\begin{equation}
\boldsymbol{E}^{(1)}(\boldsymbol{0})
=-\boldsymbol{\Sigma}_{\mathrm{E}}
\cdot\boldsymbol{E}_{\mathrm{ext}},
\label{2.18}
\end{equation}
where $\boldsymbol{\Sigma}_{\mathrm{E}}$ is the electric dipole
shielding tensor given by
\begin{equation}
\boldsymbol{\Sigma}_{\mathrm{E}}=\frac{2e^{2}}{4\pi\epsilon_{0}}
\Real\int_{\mathbb{R}^{3}}\mathrm{d}^{3}\boldsymbol{r}
\int_{\mathbb{R}^{3}}\mathrm{d}^{3}\boldsymbol{r}'\:
\Psi^{(0)\dag}(\boldsymbol{r})\frac{\boldsymbol{r}}{r^{3}}
\bar{G}\mbox{}^{(0)}(\boldsymbol{r},\boldsymbol{r}')
\boldsymbol{r}'\Psi^{(0)}(\boldsymbol{r}').
\label{2.19}
\end{equation}
To evaluate $\boldsymbol{\Sigma}_{\mathrm{E}}$, we rewrite Eq.\
(\ref{2.19}) in the form
\begin{equation}
\boldsymbol{\Sigma}_{\mathrm{E}}=\frac{2e^{2}}{4\pi\epsilon_{0}}
\Real\sum_{n,n'=-1}^{+1}
\boldsymbol{e}_{n}\boldsymbol{e}_{n'}^{*}
\int_{\mathbb{R}^{3}}\mathrm{d}^{3}\boldsymbol{r}
\int_{\mathbb{R}^{3}}\mathrm{d}^{3}\boldsymbol{r}'\:
\Psi^{(0)\dag}(\boldsymbol{r})r^{-2}
\boldsymbol{e}_{n}^{*}\cdot\boldsymbol{n}_{r}
\bar{G}\mbox{}^{(0)}(\boldsymbol{r},\boldsymbol{r}')
r'\boldsymbol{e}_{n'}\cdot\boldsymbol{n}_{r}^{\prime}
\Psi^{(0)}(\boldsymbol{r}'),
\label{2.20}
\end{equation}
where $\boldsymbol{e}_{n}$, with $n=0,\pm1$, are the unit vectors of
the cyclic basis, related to the Cartesian unit vectors through
\begin{equation}
\boldsymbol{e}_{0}=\boldsymbol{e}_{z},
\qquad
\boldsymbol{e}_{\pm1}=\mp\frac{1}{\sqrt{2}}
(\boldsymbol{e}_{x}\pm\mathrm{i}\boldsymbol{e}_{y}).
\label{2.21}
\end{equation}
In the next step, we substitute $\Psi^{(0)}(\boldsymbol{r})$, as
given by Eqs.\ (\ref{2.2}) and (\ref{2.4}), and the multipole
expansion of the generalized Dirac--Coulomb Green function, which is
\begin{eqnarray}
\bar{G}\mbox{}^{(0)}(\boldsymbol{r},\boldsymbol{r}')
&=& \frac{4\pi\epsilon_{0}}{e^{2}}
\sum_{\substack{K=-\infty \\ (K\neq0)}}^{\infty}
\sum_{M=-|K|+1/2}^{|K|-1/2}\frac{1}{rr'}
\nonumber \\
&& \times\left(
\begin{array}{cc}
\bar{g}\mbox{}^{(0)}_{K,(++)}(r,r')
\Omega_{KM}(\boldsymbol{n}_{r})
\Omega_{KM}^{\dag}(\boldsymbol{n}_{r}^{\prime}) &
-\mathrm{i}\bar{g}\mbox{}^{(0)}_{K,(+-)}(r,r')
\Omega_{KM}(\boldsymbol{n}_{r})
\Omega_{-KM}^{\dag}(\boldsymbol{n}_{r}^{\prime}) \\
\mathrm{i}\bar{g}\mbox{}^{(0)}_{K,(-+)}(r,r')
\Omega_{-KM}(\boldsymbol{n}_{r})
\Omega_{KM}^{\dag}(\boldsymbol{n}_{r}^{\prime}) &
\bar{g}\mbox{}^{(0)}_{K,(--)}(r,r')
\Omega_{-KM}(\boldsymbol{n}_{r})
\Omega_{-KM}^{\dag}(\boldsymbol{n}_{r}^{\prime})
\end{array}
\right),
\nonumber \\
&&
\label{2.22}
\end{eqnarray}
into Eq.\ (\ref{2.20}). After the angular integrals appearing in the
resulting series representation of $\boldsymbol{\Sigma}_{\mathrm{E}}$
are evaluated with the aid of the following identities obeyed by the
spherical spinors \cite{Szmy07}:
\begin{eqnarray}
\boldsymbol{e}_{0}\cdot\boldsymbol{n}_{r}\,
\Omega_{\kappa\mu}(\boldsymbol{n}_{r})
&=& -\,\frac{2\mu}{4\kappa^{2}-1}
\Omega_{-\kappa\mu}(\boldsymbol{n}_{r})
+\frac{\sqrt{(\kappa+\frac{1}{2})^{2}-\mu^{2}}}{|2\kappa+1|}
\Omega_{\kappa+1,\mu}(\boldsymbol{n}_{r})
\nonumber \\
&& +\,\frac{\sqrt{(\kappa-\frac{1}{2})^{2}-\mu^{2}}}{|2\kappa-1|}
\Omega_{\kappa-1,\mu}(\boldsymbol{n}_{r}),
\label{2.23}
\end{eqnarray}
\begin{eqnarray}
\boldsymbol{e}_{\pm1}\cdot\boldsymbol{n}_{r}\,
\Omega_{\kappa\mu}(\boldsymbol{n}_{r})
&=& \pm\,\sqrt{2}\frac{\sqrt{\kappa^{2}-(\mu\pm\frac{1}{2})^{2}}}
{4\kappa^{2}-1}\Omega_{-\kappa,\mu\pm1}(\boldsymbol{n}_{r})
\nonumber \\
&& +\,\frac{\sqrt{(\kappa\pm\mu+\frac{1}{2})
(\kappa\pm\mu+\frac{3}{2})}}{\sqrt{2}(2\kappa+1)}
\Omega_{\kappa+1,\mu\pm1}(\boldsymbol{n}_{r})
\nonumber \\
&& -\,\frac{\sqrt{(\kappa\mp\mu-\frac{1}{2})
(\kappa\mp\mu-\frac{3}{2})}}{\sqrt{2}(2\kappa-1)}
\Omega_{\kappa-1,\mu\pm1}(\boldsymbol{n}_{r}),
\label{2.24}
\end{eqnarray}
we arrive at the conclusion that for arbitrary (up to the
normalization constraint (\ref{2.3})) values of the coefficients
$a_{\pm1/2}^{(0)}$ the electric shielding tensor for the hydrogenic
ground state is a multiple of the unit dyad,
\begin{equation}
\boldsymbol{\Sigma}_{\mathrm{E}}
=\sigma_{\mathrm{E}}\boldsymbol{\mathsf{I}},
\label{2.25}
\end{equation}
and that the only contributions to the shielding constant
$\sigma_{\mathrm{E}}$ come from the terms with $K=+1$ and $K=-2$. One
has
\begin{equation}
\sigma_{\mathrm{E}}=\sigma_{\mathrm{E},1}+\sigma_{\mathrm{E},-2},
\label{2.26}
\end{equation}
where
\begin{equation}
\sigma_{\mathrm{E},1}=\frac{2}{9}
\int_{0}^{\infty}\mathrm{d}r\int_{0}^{\infty}\mathrm{d}r'\:
\left(
\begin{array}{cc}
P^{(0)}(r) & Q^{(0)}(r)
\end{array}
\right)
r^{-2}\bar{\mathsf{G}}\mbox{}^{(0)}_{1}(r,r')r'
\left(
\begin{array}{c}
P^{(0)}(r') \\
Q^{(0)}(r')
\end{array}
\right)
\label{2.27}
\end{equation}
and
\begin{equation}
\sigma_{\mathrm{E},-2}=\frac{4}{9}
\int_{0}^{\infty}\mathrm{d}r\int_{0}^{\infty}\mathrm{d}r'\:
\left(
\begin{array}{cc}
P^{(0)}(r) & Q^{(0)}(r)
\end{array}
\right)
r^{-2}\bar{\mathsf{G}}\mbox{}^{(0)}_{-2}(r,r')r'
\left(
\begin{array}{c}
P^{(0)}(r') \\
Q^{(0)}(r')
\end{array}
\right),
\label{2.28}
\end{equation}
with
\begin{equation}
\bar{\mathsf{G}}\mbox{}^{(0)}_{K}(r,r')
=\left(
\begin{array}{cc}
\bar{g}^{(0)}\mbox{}_{K,(++)}(r,r') &
\bar{g}^{(0)}\mbox{}_{K,(+-)}(r,r') \\
\bar{g}^{(0)}\mbox{}_{K,(-+)}(r,r') &
\bar{g}^{(0)}\mbox{}_{K,(--)}(r,r')
\end{array}
\right)
\label{2.29}
\end{equation}
being the radial generalized Dirac--Coulomb Green function associated
with the hydrogenic ground-state energy level.

To evaluate the integrals in Eqs.\ (\ref{2.27}) and (\ref{2.28}), we
shall exploit the following separable expansion of 
$\bar{\mathsf{G}}\mbox{}^{(0)}_{K}(r,r')$: 
\begin{equation}
\bar{\mathsf{G}}\mbox{}^{(0)}_{K}(r,r')
=\sum_{n=-\infty}^{\infty}
\frac{1}{\mu_{nK}^{(0)}-1}
\left(
\begin{array}{c}
S_{nK}^{(0)}(r) \\
T_{nK}^{(0)}(r)
\end{array}
\right)
\left(
\begin{array}{cc}
\mu_{nK}^{(0)}S_{nK}^{(0)}(r') & T_{nK}^{(0)}(r')
\end{array}
\right)
\qquad (K\neq-1),
\label{2.30}
\end{equation}
which is a corollary from the theory of the Dirac--Coulomb Sturmian
functions \cite{Szmy97}. Here
\begin{eqnarray}
S_{nK}^{(0)}(r)
&=& \sqrt{\frac{(1+\gamma_{1})(|n|+2\gamma_{K})|n|!}
{2ZN_{nK}(N_{nK}-K)\Gamma(|n|+2\gamma_{K})}}
\nonumber \\
&& \times\left(\frac{2Zr}{a_{0}}\right)^{\gamma_{K}}
\textrm{e}^{-Zr/a_{0}}
\left[L_{|n|-1}^{(2\gamma_{K})}\left(\frac{2Zr}{a_{0}}\right)
+\frac{K-N_{nK}}{|n|+2\gamma_{K}}
L_{|n|}^{(2\gamma_{K})}\left(\frac{2Zr}{a_{0}}\right)\right]
\label{2.31}
\end{eqnarray}
and
\begin{eqnarray}
T_{nK}^{(0)}(r)
&=& \sqrt{\frac{(1-\gamma_{1})(|n|+2\gamma_{K})|n|!}
{2ZN_{nK}(N_{nK}-K)\Gamma(|n|+2\gamma_{K})}}
\nonumber \\
\qquad
&& \times\left(\frac{2Zr}{a_{0}}\right)^{\gamma_{K}}
\textrm{e}^{-Zr/a_{0}}
\left[L_{|n|-1}^{(2\gamma_{K})}\left(\frac{2Zr}{a_{0}}\right)-
\frac{K-N_{nK}}{|n|+2\gamma_{K}}
L_{|n|}^{(2\gamma_{K})}\left(\frac{2Zr}{a_{0}}\right)\right]
\label{2.32}
\end{eqnarray}
(with $L_{n}^{(\alpha)}(\rho)$ denoting the generalized Laguerre
polynomials \cite{Magn66}; we define
$L_{-1}^{(\alpha)}(\rho)\equiv0$) are the radial Dirac--Coulomb
Sturmian functions associated with the hydrogenic ground-state energy
level, and
\begin{equation}
\mu_{nK}^{(0)}=\frac{|n|+\gamma_{K}+N_{nK}}{\gamma_{1}+1},
\label{2.33}
\end{equation}
where
\begin{equation}
N_{nK}=\pm\sqrt{(|n|+\gamma_{K})^{2}+(Z\alpha)^{2}}
=\pm\sqrt{|n|^{2}+2|n|\gamma_{K}+K^{2}}
\label{2.34}
\end{equation}
is the `apparent principal quantum number' (notice that it may assume
positive as well as negative values!). The following sign convention
applies to the definition (\ref{2.34}): the plus sign should be
chosen for $n>0$ and the minus one for $n<0$; for $n=0$ one chooses
the plus sign if $K<0$ and the minus sign if $K>0$.

At first, we attack the double integral in the expression for
$\sigma_{\mathrm{E},1}$. Inserting Eqs.\ (\ref{2.5}), (\ref{2.6}) and
(\ref{2.30})--(\ref{2.33}) into the right-hand side of Eq.\
(\ref{2.27}), taking the two resulting separated integrals with
the aid of the known formula \cite[Eq.\ (7.414.11)]{Grad94}
\begin{equation}
\int_{0}^{\infty}\mathrm{d}\rho\:\rho^{\beta}\mathrm{e}^{-\rho}
L_{n}^{(\alpha)}(\rho)
=\frac{\Gamma(\beta+1)\Gamma(n+\alpha-\beta)}{n!\Gamma(\alpha-\beta)}
=(-)^{n}\frac{\Gamma(\beta+1)\Gamma(\beta-\alpha+1)}
{n!\Gamma(\beta-\alpha-n+1)}
\qquad 
[\Real(\beta)>-1]
\label{2.35}
\end{equation}
and employing Eq.\ (\ref{2.34}), we obtain $\sigma_{\mathrm{E},1}$ in
the form of the following finite sum:
\begin{eqnarray}
\sigma_{\mathrm{E},1} &=& \frac{2}{9Z}
(\gamma_{1}+1)^{2}(2\gamma_{1}+1)\Gamma(2\gamma_{1}-1)
\nonumber \\
&& \times \sum_{n=-2}^{2}(-)^{n}
\frac{(N_{n1}-1)(|n|+N_{n1}+1)(|n|-\gamma_{1}N_{n1}-\gamma_{1}+1)}
{N_{n1}(2-|n|)!\Gamma(|n|+2\gamma_{1}+1)(|n|+N_{n1}-1)}.
\label{2.36}
\end{eqnarray}
Using again Eq.\ (\ref{2.34}), after some algebra, the right-hand
side of Eq.\ (\ref{2.36}) simplifies considerably, yielding
\begin{equation}
\sigma_{\mathrm{E},1}=\frac{2\gamma_{1}+1}{9Z}.
\label{2.37}
\end{equation}
As one of the two separated integrals evaluated above converges at
its lower integration limit provided $\gamma_{1}>1/2$, we have the
following constraint on the nuclear charge:
\begin{equation}
Z<\alpha^{-1}\frac{\sqrt{3}}{2}\simeq118.67.
\label{2.38}
\end{equation}

Evaluation of the double integral on the right-hand side of Eq.\
(\ref{2.28}) appears to be much more cumbersome. Proceeding initially
as in the case discussed above, employing Eq.\ (\ref{2.34}) and the
trivial but useful identity
\begin{equation}
\gamma_{2}^{2}=\gamma_{1}^{2}+3,
\label{2.39}
\end{equation}
with much labor we obtain
\begin{eqnarray}
\sigma_{\mathrm{E},-2}
&=& \frac{1}{2Z}\frac{(\gamma_{1}+1)\Gamma(\gamma_{2}+\gamma_{1}-1)
\Gamma(\gamma_{2}+\gamma_{1}+2)}{\Gamma(\gamma_{2}-\gamma_{1}-1)
\Gamma(\gamma_{2}-\gamma_{1}+2)\Gamma(2\gamma_{1}+1)}
\nonumber \\
&&
\times\sum_{n=-\infty}^{\infty}
\frac{\Gamma(|n|+\gamma_{2}-\gamma_{1}+1)
\Gamma(|n|+\gamma_{2}-\gamma_{1}-2)}{|n|!\Gamma(|n|+2\gamma_{2}+1)}
\frac{N_{n,-2}+2}{N_{n,-2}}
\nonumber \\
&& \times\frac{(|n|+\gamma_{2}+\gamma_{1}+1-\gamma_{1}N_{n,-2})
[(\gamma_{1}-1)(2\gamma_{1}+5)-(2\gamma_{1}-1)
(|n|+\gamma_{2}+N_{n,-2})]}
{|n|+\gamma_{2}-\gamma_{1}-1+N_{n,-2}}.
\nonumber \\
&&
\label{2.40}
\end{eqnarray}
Collecting together those terms in the above series which correspond
to the same absolute value of $n$, again with the use of Eqs.\
(\ref{2.34}) and (\ref{2.39}), we arrive at
\begin{eqnarray}
\sigma_{\mathrm{E},-2}
&=& -\,\frac{4}{9Z}
\frac{(\gamma_{1}+1)\Gamma(\gamma_{2}+\gamma_{1}-1)
\Gamma(\gamma_{2}+\gamma_{1}+2)}{\Gamma(\gamma_{2}-\gamma_{1}-1)
\Gamma(\gamma_{2}-\gamma_{1}+2)\Gamma(2\gamma_{1}+1)}
\nonumber \\
&& \times\sum_{n=0}^{\infty}\frac{\Gamma(n+\gamma_{2}-\gamma_{1}-2)
\Gamma(n+\gamma_{2}-\gamma_{1})}{n!\Gamma(n+2\gamma_{2}+1)}
\nonumber \\
&& \quad
\times\big[3(\gamma_{1}-1)(n+\gamma_{2}-\gamma_{1}-2)
(n+\gamma_{2}-\gamma_{1})+(2\gamma_{1}^{2}+2\gamma_{1}-3)
(n+\gamma_{2}-\gamma_{1})-3(\gamma_{1}-2)\big].
\nonumber \\
&&
\label{2.41}
\end{eqnarray}
The series in Eq.\ (\ref{2.41}) does not terminate. Nevertheless, we
shall show that $\sigma_{\mathrm{E},-2}$ may be expressed in the form
which is as elementary as the one in the case of
$\sigma_{\mathrm{E},1}$. To this end, we invoke the relationship
\begin{equation}
\sum_{n=0}^{\infty}\frac{\Gamma(n+a_{1})\Gamma(n+a_{2})}
{\Gamma(n+b)}\frac{z^{n}}{n!}
=\frac{\Gamma(a_{1})\Gamma(a_{2})}{\Gamma(b)}
\,{}_{2}F_{1}
\left(
\begin{array}{c}
a_{1}, a_{2} \\
b
\end{array}
;z
\right)
\qquad (|z|\leqslant1),
\label{2.42}
\end{equation}
where ${}_{2}F_{1}$ is the hypergeometric function. Transforming Eq.\
(\ref{2.41}) with the aid of the above formula results in
\begin{eqnarray}
\sigma_{\mathrm{E},-2}
&=& -\,\frac{4}{9Z}
\frac{(\gamma_{1}+1)\Gamma(\gamma_{2}+\gamma_{1}-1)
\Gamma(\gamma_{2}+\gamma_{1}+2)}{(\gamma_{2}-\gamma_{1}-2)
(\gamma_{2}-\gamma_{1})(\gamma_{2}-\gamma_{1}+1)\Gamma(2\gamma_{1}+1)
\Gamma(2\gamma_{2}+1)}
\nonumber \\
&& \times\left[3(\gamma_{1}-1)(\gamma_{2}-\gamma_{1}-2)
(\gamma_{2}-\gamma_{1})
\,{}_{2}F_{1}
\left(
\begin{array}{c}
\gamma_{2}-\gamma_{1}-1,\gamma_{2}-\gamma_{1}+1 \\
2\gamma_{2}+1
\end{array}
;1
\right)
\right.
\nonumber \\
&& \quad 
+\,(2\gamma_{1}^{2}+2\gamma_{1}-3)(\gamma_{2}-\gamma_{1})
\,{}_{2}F_{1}
\left(
\begin{array}{c}
\gamma_{2}-\gamma_{1}-2,\gamma_{2}-\gamma_{1}+1 \\
2\gamma_{2}+1
\end{array}
;1
\right)
\nonumber \\
&& \quad
\left.-\,3(\gamma_{1}-2)
\,{}_{2}F_{1}
\left(
\begin{array}{c}
\gamma_{2}-\gamma_{1}-2,\gamma_{2}-\gamma_{1} \\
2\gamma_{2}+1
\end{array}
;1
\right)
\right].
\label{2.43}
\end{eqnarray}
Now, we owe to Gauss the following identity \cite{Magn66}:
\begin{equation}
{}_{2}F_{1}
\left(
\begin{array}{c}
a_{1}, a_{2} \\
b
\end{array}
;1
\right)
=\frac{\Gamma(b)\Gamma(b-a_{1}-a_{2})}
{\Gamma(b-a_{1})\Gamma(b-a_{2})}
\qquad [\Real(b-a_{1}-a_{2})>0].
\label{2.44}
\end{equation}
Applying it to the three ${}_{2}F_{1}$ functions appearing in Eq.\
(\ref{2.43}), we find
\begin{equation}
{}_{2}F_{1}
\left(
\begin{array}{c}
\gamma_{2}-\gamma_{1}-1,\gamma_{2}-\gamma_{1}+1 \\
2\gamma_{2}+1
\end{array}
;1
\right)
=\frac{\Gamma(2\gamma_{1}+1)\Gamma(2\gamma_{2}+1)}
{\Gamma(\gamma_{2}+\gamma_{1})\Gamma(\gamma_{2}+\gamma_{1}+2)},
\label{2.45}
\end{equation}
\begin{equation}
{}_{2}F_{1}
\left(
\begin{array}{c}
\gamma_{2}-\gamma_{1}-2,\gamma_{2}-\gamma_{1}+1 \\
2\gamma_{2}+1
\end{array}
;1
\right)
=\frac{\Gamma(2\gamma_{1}+2)\Gamma(2\gamma_{2}+1)}
{\Gamma(\gamma_{2}+\gamma_{1})\Gamma(\gamma_{2}+\gamma_{1}+3)}
\label{2.46}
\end{equation}
and
\begin{equation}
{}_{2}F_{1}
\left(
\begin{array}{c}
\gamma_{2}-\gamma_{1}-2,\gamma_{2}-\gamma_{1} \\
2\gamma_{2}+1
\end{array}
;1
\right)
=\frac{\Gamma(2\gamma_{1}+3)\Gamma(2\gamma_{2}+1)}
{\Gamma(\gamma_{2}+\gamma_{1}+1)\Gamma(\gamma_{2}+\gamma_{1}+3)}.
\label{2.47}
\end{equation}
Plugging Eqs.\ (\ref{2.45})--(\ref{2.47}) into Eq.\ (\ref{2.43}),
after some further simplifications based on the use of the relation
(\ref{2.39}), we arrive at the afore-announced elementary
representation of the $K=-2$ component of the shielding constant:
\begin{equation}
\sigma_{\mathrm{E},-2}=-\frac{2(\gamma_{1}-4)}{9Z}.
\label{2.48}
\end{equation}

Insertion of $\sigma_{\mathrm{E},1}$ and $\sigma_{\mathrm{E},-2}$ as
given in Eqs.\ (\ref{2.37}) and (\ref{2.48}), respectively, into Eq.\
(\ref{2.26}) leads us to the conclusion that the electric dipole
shielding constant for the Dirac hydrogenic atom in its ground state
is
\begin{equation}
\sigma_{\mathrm{E}}=Z^{-1},
\label{2.49}
\end{equation}
as it should be.

In the nonrelativistic limit 
\begin{equation}
\gamma_{1}\simeq1-\frac{1}{2}(Z\alpha)^{2},
\label{2.50}
\end{equation}
so that one has
\begin{equation}
\sigma_{\mathrm{E},1}\simeq\frac{1}{3Z}
\left[1-\frac{1}{3}(Z\alpha)^{2}\right]
\label{2.51}
\end{equation}
and
\begin{equation}
\sigma_{\mathrm{E},-2}\simeq\frac{2}{3Z}
\left[1+\frac{1}{6}(Z\alpha)^{2}\right].
\label{2.52}
\end{equation}
%
%
%\newpage
%
\section{The magnetic dipole shielding constant}
\label{III}
\setcounter{equation}{0}
We proceed to the evaluation of the magnetic dipole shielding
constant. The model of the hydrogen-like atom we shall adopt for this
purpose is the same as in the preceding section.

In a weak, constant, uniform magnetic field
$\boldsymbol{B}_{\mathrm{ext}}$, the atomic ground energy level
$\mathcal{E}^{(0)}$ splits into two, their energies being given, to
the first order in $\boldsymbol{B}_{\mathrm{ext}}$, by
\begin{equation}
\mathcal{E}_{\mu}\simeq\mathcal{E}^{(0)}+\mathcal{E}_{\mu}^{(1)}
\qquad ({\textstyle\mu=\pm\frac{1}{2}}),
\label{3.1}
\end{equation}
with
\begin{equation}
\mathcal{E}_{\mu}^{(1)}=\sgn(\mu)\frac{2\gamma_{1}+1}{3}
\mu_{\mathrm{B}}B_{\mathrm{ext}},
\label{3.2}
\end{equation}
where $\mu_{B}$ is the Bohr magneton. The corresponding wave
functions, to the same approximation order, are
\begin{equation}
\Psi_{\mu}(\boldsymbol{r})\simeq\Psi_{\mu}^{(0)}(\boldsymbol{r})
+\Psi_{\mu}^{(1)}(\boldsymbol{r})
\qquad ({\textstyle\mu=\pm\frac{1}{2}}),
\label{3.3}
\end{equation}
with $\Psi_{\mu}^{(0)}(\boldsymbol{r})$ given by Eq.\ (\ref{2.4})
(the space quantization axis being chosen along the external magnetic
field direction) and with
\begin{equation}
\Psi_{\mu}^{(1)}(\boldsymbol{r})
=-\frac{1}{2}ec\boldsymbol{B}_{\mathrm{ext}}
\cdot\int_{\mathbb{R}^{3}}\mathrm{d}^{3}\boldsymbol{r}'\:
\bar{G}\mbox{}^{(0)}(\boldsymbol{r},\boldsymbol{r}')
(\boldsymbol{r}'\times\boldsymbol{\alpha})
\Psi_{\mu}^{(0)}(\boldsymbol{r}').
\label{3.4}
\end{equation}
Here, $\boldsymbol{\alpha}$ is the Dirac $4\times4$ vector matrix
given standardly by
\begin{equation}
\boldsymbol{\alpha}
=\left(
\begin{array}{cc}
\boldsymbol{0} & \boldsymbol{\sigma} \\
\boldsymbol{\sigma} & \boldsymbol{0}
\end{array}
\right),
\label{3.5}
\end{equation}
where $\boldsymbol{\sigma}$ is the vector composed of the Pauli
matrices.

According to the Dirac theory, there is an electric current with
density
\begin{equation}
\boldsymbol{j}_{\mu}(\boldsymbol{r})
=-ec\Psi_{\mu}^{\dag}(\boldsymbol{r})\boldsymbol{\alpha}
\Psi_{\mu}(\boldsymbol{r})
\label{3.6}
\end{equation}
associated with the electron being in the state
$\Psi_{\mu}(\boldsymbol{r})$. Using the Biot--Savart law, we find
that at the point $\boldsymbol{r}$ the magnetic field due to the
current distribution $\boldsymbol{j}_{\mu}(\boldsymbol{r}')$ is
\begin{equation}
\boldsymbol{B}_{\mu}(\boldsymbol{r})=\frac{\mu_{0}}{4\pi}
\int_{\mathbb{R}^{3}}\mathrm{d}^{3}\boldsymbol{r}'\:
\boldsymbol{j}_{\mu}(\boldsymbol{r}')
\times\frac{\boldsymbol{r}-\boldsymbol{r}'}
{|\boldsymbol{r}-\boldsymbol{r}'|^{3}}.
\label{3.7}
\end{equation}
In the particular case when the observation point is located at the
nucleus, i.e., at $\boldsymbol{r}=\boldsymbol{0}$, the above
expression simplifies to
\begin{equation}
\boldsymbol{B}_{\mu}(\boldsymbol{0})=\frac{\mu_{0}}{4\pi}
\int_{\mathbb{R}^{3}}\mathrm{d}^{3}\boldsymbol{r}'\:
\frac{\boldsymbol{r}'\times\boldsymbol{j}_{\mu}(\boldsymbol{r}')}
{r^{\prime\,3}}.
\label{3.8}
\end{equation}
If, as we have assumed above, the electronic wave function
$\Psi_{\mu}(\boldsymbol{r})$ is known to the first order in the
perturbing field $\boldsymbol{B}_{\mathrm{ext}}$, the current
$\boldsymbol{j}_{\mu}(\boldsymbol{r})$ may be approximated as
\begin{equation}
\boldsymbol{j}_{\mu}(\boldsymbol{r})
\simeq\boldsymbol{j}_{\mu}^{(0)}(\boldsymbol{r})
+\boldsymbol{j}_{\mu}^{(1)}(\boldsymbol{r}),
\label{3.9}
\end{equation}
with
\begin{equation}
\boldsymbol{j}_{\mu}^{(0)}(\boldsymbol{r})
=-ec\Psi_{\mu}^{(0)\dag}(\boldsymbol{r})\boldsymbol{\alpha}
\Psi_{\mu}^{(0)}(\boldsymbol{r})
\label{3.10}
\end{equation}
and
\begin{equation}
\boldsymbol{j}_{\mu}^{(1)}(\boldsymbol{r})
=-2ec\Real[\Psi_{\mu}^{(0)\dag}(\boldsymbol{r})\boldsymbol{\alpha}
\Psi_{\mu}^{(1)}(\boldsymbol{r})].
\label{3.11}
\end{equation}
Consequently, for the magnetic field at the nucleus location we have
\begin{equation}
\boldsymbol{B}_{\mu}(\boldsymbol{0})
\simeq\boldsymbol{B}_{\mu}^{(0)}(\boldsymbol{0})
+\boldsymbol{B}_{\mu}^{(1)}(\boldsymbol{0}),
\label{3.12}
\end{equation}
with
\begin{equation}
\boldsymbol{B}_{\mu}^{(0)}(\boldsymbol{0})=-ec\,\frac{\mu_{0}}{4\pi}
\int_{\mathbb{R}^{3}}\mathrm{d}^{3}\boldsymbol{r}'\:
\frac{\Psi_{\mu}^{(0)\dag}(\boldsymbol{r}')
\boldsymbol{r}'\times\boldsymbol{\alpha}
\Psi_{\mu}^{(0)}(\boldsymbol{r}')}{r^{\prime\,3}}
\label{3.13}
\end{equation}
and
\begin{equation}
\boldsymbol{B}_{\mu}^{(1)}(\boldsymbol{0})=-2ec\,\frac{\mu_{0}}{4\pi}
\Real\int_{\mathbb{R}^{3}}\mathrm{d}^{3}\boldsymbol{r}'\:
\frac{\Psi_{\mu}^{(0)\dag}(\boldsymbol{r}')
\boldsymbol{r}'\times\boldsymbol{\alpha}
\Psi_{\mu}^{(1)}(\boldsymbol{r}')}{r^{\prime\,3}}.
\label{3.14}
\end{equation}
If Eq.\ (\ref{3.13}) is rewritten in the form
\begin{equation}
\boldsymbol{B}_{\mu}^{(0)}(\boldsymbol{0})=-ec\,\frac{\mu_{0}}{4\pi}
\sum_{n=-1}^{1}\boldsymbol{e}_{n}^{*}
\int_{\mathbb{R}^{3}}\mathrm{d}^{3}\boldsymbol{r}'\:
\frac{\Psi_{\mu}^{(0)\dag}(\boldsymbol{r}')r'\boldsymbol{e}_{n}
\cdot(\boldsymbol{n}_{r}^{\prime}\times\boldsymbol{\alpha})
\Psi_{\mu}^{(0)}(\boldsymbol{r}')}{r^{\prime\,3}},
\label{3.15}
\end{equation}
using Eqs.\ (\ref{2.4})--(\ref{2.6}) and the spherical spinor
identities \cite{Szmy07}
\begin{eqnarray}
\boldsymbol{e}_{0}
\cdot(\boldsymbol{n}_{r}\times\boldsymbol{\sigma})\,
\Omega_{\kappa\mu}(\boldsymbol{n}_{r})
&=& \mathrm{i}\frac{4\mu\kappa}{4\kappa^{2}-1}
\Omega_{-\kappa\mu}(\boldsymbol{n}_{r})
+\mathrm{i}\frac{\sqrt{(\kappa+\frac{1}{2})^{2}-\mu^{2}}}{|2\kappa+1|}
\Omega_{\kappa+1,\mu}(\boldsymbol{n}_{r})
\nonumber \\
&& -\,\mathrm{i}
\frac{\sqrt{(\kappa-\frac{1}{2})^{2}-\mu^{2}}}{|2\kappa-1|}
\Omega_{\kappa-1,\mu}(\boldsymbol{n}_{r}),
\label{3.16}
\end{eqnarray}
\begin{eqnarray}
\boldsymbol{e}_{\pm1}
\cdot(\boldsymbol{n}_{r}\times\boldsymbol{\sigma})\,
\Omega_{\kappa\mu}(\boldsymbol{n}_{r})
&=& \mp \mathrm{i}2\sqrt{2}\kappa
\frac{\sqrt{\kappa^{2}-(\mu\pm\frac{1}{2})^{2}}}{4\kappa^{2}-1}
\Omega_{-\kappa,\mu\pm1}(\boldsymbol{n}_{r})
\nonumber \\
&& +\,\mathrm{i}\frac{\sqrt{(\kappa\pm\mu+\frac{1}{2})
(\kappa\pm\mu+\frac{3}{2})}}{\sqrt{2}(2\kappa+1)}
\Omega_{\kappa+1,\mu\pm1}(\boldsymbol{n}_{r})
\nonumber \\
&& +\,\mathrm{i}\frac{\sqrt{(\kappa\mp\mu-\frac{1}{2})
(\kappa\mp\mu-\frac{3}{2})}}{\sqrt{2}(2\kappa-1)}
\Omega_{\kappa-1,\mu\pm1}(\boldsymbol{n}_{r}),
\label{3.17}
\end{eqnarray}
with no difficulty one finds that
\begin{equation}
\boldsymbol{B}_{\mu}^{(0)}(\boldsymbol{0})
=\sgn(-\mu)\frac{8}{3\gamma_{1}(2\gamma_{1}-1)}
b_{0}\boldsymbol{n}_{\mathrm{ext}},
\label{3.18}
\end{equation}
where
\begin{equation}
b_{0}=\frac{\mu_{0}}{4\pi}\frac{\mu_{\mathrm{B}}}{a_{0}^{3}}
\label{3.19}
\end{equation}
is the atomic unit of the magnetic field induction, while
$\boldsymbol{n}_{\mathrm{ext}}$ is the unit vector along
$\boldsymbol{B}_{\mathrm{ext}}$. The radial integral 
\begin{displaymath}
\int_{0}^{\infty}\mathrm{d}r\:r^{-2}P^{(0)}(r)Q^{(0)}(r),
\end{displaymath}
encountered in the course of evaluation of
$\boldsymbol{B}_{\mu}^{(0)}(\boldsymbol{0})$, converges at its lower
limit provided $\gamma_{1}>1/2$, which constrains the nuclear charge
as in Eq.\ (\ref{2.38}).

To evaluate the first order approximation to the magnetic field
induced at the nucleus, we plug Eq.\ (\ref{3.4}) into Eq.\
(\ref{3.14}). This gives
\begin{equation}
\boldsymbol{B}_{\mu}^{(1)}(\boldsymbol{0})
=-\boldsymbol{\Sigma}_{\mathrm{M},\mu}
\cdot\boldsymbol{B}_{\mathrm{ext}},
\label{3.20}
\end{equation}
where
\begin{equation}
\boldsymbol{\Sigma}_{\mathrm{M},\mu}
=-\frac{e^{2}}{4\pi\epsilon_{0}}
\Real\int_{\mathbb{R}^{3}}\mathrm{d}^{3}\boldsymbol{r}
\int_{\mathbb{R}^{3}}\mathrm{d}^{3}\boldsymbol{r}'\:
\Psi_{\mu}^{(0)\dag}(\boldsymbol{r})
\frac{\boldsymbol{r}\times\boldsymbol{\alpha}}{r^{3}}
\bar{G}\mbox{}^{(0)}(\boldsymbol{r},\boldsymbol{r}')
\boldsymbol{r}'\times\boldsymbol{\alpha}
\Psi_{\mu}^{(0)}(\boldsymbol{r}')
\label{3.21}
\end{equation}
is the magnetic dipole shielding tensor. Rewriting Eq.\ (\ref{3.21})
in the form
{\small
\begin{equation}
\boldsymbol{\Sigma}_{\mathrm{M},\mu}
=-\frac{e^{2}}{4\pi\epsilon_{0}}\Real\sum_{n,n'=-1}^{+1}
\boldsymbol{e}_{n}\boldsymbol{e}_{n'}^{*}
\int_{\mathbb{R}^{3}}\mathrm{d}^{3}\boldsymbol{r}
\int_{\mathbb{R}^{3}}\mathrm{d}^{3}\boldsymbol{r}'\:
\Psi_{\mu}^{(0)\dag}(\boldsymbol{r})r^{-2}
\boldsymbol{e}_{n}^{*}\cdot(\boldsymbol{n}_{r}
\times\boldsymbol{\alpha})
\bar{G}\mbox{}^{(0)}(\boldsymbol{r},\boldsymbol{r}')
r'\boldsymbol{e}_{n'}\cdot(\boldsymbol{n}_{r}^{\prime}
\times\boldsymbol{\alpha})\Psi_{\mu}^{(0)}(\boldsymbol{r}'),
\label{3.22}
\end{equation}
}
inserting the multipole representation (\ref{2.22}) of the
generalized Dirac--Coulomb Green function into Eq.\ (\ref{3.22}) and
using the relations (\ref{3.16}) and (\ref{3.17}) to carry out
integrations over the angles of the vectors $\boldsymbol{r}$ and
$\boldsymbol{r}'$, after some labor we discover that, like its
electric counterpart, the tensor 
$\boldsymbol{\Sigma}_{\mathrm{M},\mu}$ is a multiple of the unit
dyad:
\begin{equation}
\boldsymbol{\Sigma}_{\mathrm{M},\mu}
=\sigma_{\mathrm{M}}\mathsf{I}.
\label{3.23}
\end{equation}
As the notation used indicates, the factor $\sigma_{\mathrm{M}}$,
being the magnetic shielding constant for the system under study, is
independent of the angular momentum projection quantum number $\mu$;
it is given by the sum
\begin{equation}
\sigma_{\mathrm{M}}=\sigma_{\mathrm{M},-1}
+\sigma_{\mathrm{M},2},
\label{3.24}
\end{equation}
where
\begin{equation}
\sigma_{\mathrm{M},-1}
=-\frac{4}{9}\int_{0}^{\infty}\mathrm{d}r
\int_{0}^{\infty}\mathrm{d}r'\:
\left(
\begin{array}{cc}
Q^{(0)}(r) & P^{(0)}(r)
\end{array}
\right)
r^{-2}\bar{\mathsf{G}}\mbox{}^{(0)}_{-1}(r,r')r'
\left(
\begin{array}{c}
Q^{(0)}(r') \\ 
P^{(0)}(r')
\end{array}
\right)
\label{3.25}
\end{equation}
and
\begin{equation}
\sigma_{\mathrm{M},2}
=-\frac{2}{9}\int_{0}^{\infty}\mathrm{d}r
\int_{0}^{\infty}\mathrm{d}r'\:
\left(
\begin{array}{cc}
Q^{(0)}(r) & P^{(0)}(r)
\end{array}
\right)
r^{-2}\bar{\mathsf{G}}\mbox{}^{(0)}_{2}(r,r')r'
\left(
\begin{array}{c}
Q^{(0)}(r') \\ 
P^{(0)}(r')
\end{array}
\right).
\label{3.26}
\end{equation}

The Sturmian expansion of $\bar{\mathsf{G}}\mbox{}^{(0)}_{K}(r,r')$
given in Eq.\ (\ref{2.30}) is inapplicable in the case of $K=-1$ and
thus it cannot be used for the purpose of evaluation of the double
integral in Eq.\ (\ref{3.25}). The valid Sturmian representation of
$\bar{\mathsf{G}}\mbox{}^{(0)}_{-1}(r,r')$ which we employ for that
purpose is \cite{Szmy97}
\begin{eqnarray}
\bar{\mathsf{G}}\mbox{}^{(0)}_{-1}(r,r')
&=& \sum_{\scriptstyle n=-\infty\atop\scriptstyle(n\neq0)}^{\infty}
\frac{1}{\mu_{n,-1}^{(0)}-1}
\left(
\begin{array}{c}
S_{n,-1}^{(0)}(r) \\
T_{n,-1}^{(0)}(r)
\end{array}
\right)
\left(
\begin{array}{cc}
\mu_{n,-1}^{(0)}S_{n,-1}^{(0)}(r') & T_{n,-1}^{(0)}(r')
\end{array}
\right)
\nonumber \\
&& +\left(\gamma_{1}-{\textstyle\frac{1}{2}}\right)
\left(
\begin{array}{c}
S_{0,-1}^{(0)}(r) \\
T_{0,-1}^{(0)}(r)
\end{array}
\right)
\left(
\begin{array}{cc}
S_{0,-1}^{(0)}(r') & T_{0,-1}^{(0)}(r')
\end{array}
\right)
\nonumber \\
&& +\left(
\begin{array}{c}
I^{(0)}(r) \\
K^{(0)}(r)
\end{array}
\right)
\left(
\begin{array}{cc}
S_{0,-1}^{(0)}(r') & T_{0,-1}^{(0)}(r')
\end{array}
\right)
\nonumber \\
&& +\left(
\begin{array}{c}
S_{0,-1}^{(0)}(r) \\
T_{0,-1}^{(0)}(r)
\end{array}
\right)
\left(
\begin{array}{cc}
J^{(0)}(r') & K^{(0)}(r')
\end{array}
\right),
\label{3.27}
\end{eqnarray}
where
\begin{equation}
I^{(0)}(r)=\left(\gamma_{1}-{\textstyle\frac{1}{2}}\right)
S_{0,-1}^{(0)}(r)
+\gamma_{1}\left(\frac{1+\gamma_{1}}{\alpha}\frac{r}{a_{0}}
+Z\alpha\right)T_{0,-1}^{(0)}(r),
\label{3.28}
\end{equation}
\begin{eqnarray}
J^{(0)}(r) &=& I^{(0)}(r)+S_{0,-1}^{(0)}(r)
\nonumber \\
&=& \left(\gamma_{1}+{\textstyle\frac{1}{2}}\right)
S_{0,-1}^{(0)}(r)
+\gamma_{1}\left(\frac{1+\gamma_{1}}{\alpha}\frac{r}{a_{0}}
+Z\alpha\right)T_{0,-1}^{(0)}(r)
\label{3.29}
\end{eqnarray}
and
\begin{equation}
K^{(0)}(r)
=\gamma_{1}\left(\frac{1-\gamma_{1}}{\alpha}\frac{r}{a_{0}}
-Z\alpha\right)S_{0,-1}^{(0)}(r)
-\left(\gamma_{1}-{\textstyle\frac{1}{2}}\right)T_{0,-1}^{(0)}(r).
\label{3.30}
\end{equation}
Exploiting Eqs.\ (\ref{2.5}), (\ref{2.6}) and
(\ref{2.31})--(\ref{2.35}), we find that
\begin{eqnarray}
\sigma_{\mathrm{M},-1} &=& -\,\frac{8Z\alpha^{2}}{9}
(\gamma_{1}+1)(2\gamma_{1}+1)
\Gamma(2\gamma_{1}-1)
\nonumber \\
&& \times\sum_{\substack{n=-2 \\ (n\neq0)}}^{+2}(-)^{n}
\frac{N_{n,-1}+1}{N_{n,-1}}\frac{|n|+1}
{(2-|n|)!\Gamma(|n|+2\gamma_{1}+1)(|n|+N_{n,-1}-1)}
\nonumber \\
&& -\,\frac{2Z\alpha^{2}}{9}\frac{2\gamma_{1}+1}{\gamma_{1}}
-\frac{2Z\alpha^{2}}{9}\frac{2\gamma_{1}+1}{2\gamma_{1}-1}
+\frac{2Z\alpha^{2}}{9}\frac{2\gamma_{1}+1}{\gamma_{1}},
\label{3.31}
\end{eqnarray}
terms on the right-hand side being ordered in correspondence with
Eq.\ (\ref{3.27}). The first of them is readily found to be
$8Z\alpha^{2}/[9\gamma_{1}(2\gamma_{1}-1)]$, so that 
$\sigma_{\mathrm{M},-1}$ is
\begin{equation}
\sigma_{\mathrm{M},-1}=-\frac{2Z\alpha^{2}}{9}
\frac{2\gamma_{1}^{2}+\gamma_{1}-4}{\gamma_{1}(2\gamma_{1}-1)}.
\label{3.32}
\end{equation}

To calculate $\sigma_{\mathrm{M},2}$ from Eq.\ (\ref{3.26}), we use
$\bar{\mathsf{G}}\mbox{}^{(0)}_{2}(r,r')$ in the form (\ref{2.30}).
Applying Eqs.\ (\ref{2.5}), (\ref{2.6}), (\ref{2.31})--(\ref{2.35})
and (\ref{2.39}), we arrive at
\begin{eqnarray}
\sigma_{\mathrm{M},2}
&=& \frac{2Z\alpha^{2}}{9}\frac{\Gamma(\gamma_{2}+\gamma_{1}-1)
\Gamma(\gamma_{2}+\gamma_{1}+2)}
{\Gamma(\gamma_{2}-\gamma_{1}-1)\Gamma(\gamma_{2}-\gamma_{1}+2)
\Gamma(2\gamma_{1}+1)}
\nonumber \\
&& \times\sum_{n=-\infty}^{\infty}
\frac{\Gamma(|n|+\gamma_{2}-\gamma_{1}-2)
\Gamma(|n|+\gamma_{2}-\gamma_{1}+2)}{|n|!\Gamma(|n|+2\gamma_{2}+1)}
\frac{N_{n2}-2}{N_{n2}}
\nonumber \\
&& \times\frac{3|n|+3\gamma_{2}+\gamma_{1}+1+3N_{n2}}
{|n|+\gamma_{2}-\gamma_{1}-1+N_{n2}}.
\label{3.33}
\end{eqnarray}
The simplification is achieved after one collects terms with the same
absolute value of $n$. This gives
\begin{equation}
\sigma_{\mathrm{M},2}
=\frac{2Z\alpha^{2}}{9}\frac{\Gamma(\gamma_{2}+\gamma_{1}-1)
\Gamma(\gamma_{2}+\gamma_{1}+2)}
{\Gamma(\gamma_{2}-\gamma_{1}-1)\Gamma(\gamma_{2}-\gamma_{1}+2)
\Gamma(2\gamma_{1}+1)}
\sum_{n=0}^{\infty}\frac{\Gamma(n+\gamma_{2}-\gamma_{1}-1)
\Gamma(n+\gamma_{2}-\gamma_{1}+2)}{n!\Gamma(n+2\gamma_{2}+1)
(n+\gamma_{2}-\gamma_{1})},
\label{3.34}
\end{equation}
which is the final expression for $\sigma_{\mathrm{M},2}$ provided in
Ref.\ \cite{Chen09}. We shall show that, similarly to the series in
Eq.\ (\ref{2.41}), the one in Eq.\ (\ref{3.34}) may be also summed to
a closed elementary form. To this end, we rewrite the latter equation
as
\begin{eqnarray}
\sigma_{\mathrm{M},2} &=& \frac{2Z\alpha^{2}}{9}
\frac{\Gamma(\gamma_{2}+\gamma_{1}-1)\Gamma(\gamma_{2}+\gamma_{1}+2)}
{\Gamma(\gamma_{2}-\gamma_{1}-1)\Gamma(\gamma_{2}-\gamma_{1}+2)
\Gamma(2\gamma_{1}+1)}
\nonumber \\
&& \times\left[\sum_{n=0}^{\infty}
\frac{\Gamma(n+\gamma_{2}-\gamma_{1}-1)
\Gamma(n+\gamma_{2}-\gamma_{1})}{n!\Gamma(n+2\gamma_{2}+1)}\right.
\nonumber \\
&& \quad
\left.+\sum_{n=0}^{\infty}\frac{\Gamma(n+\gamma_{2}-\gamma_{1}-1)
\Gamma(n+\gamma_{2}-\gamma_{1}+1)}{n!\Gamma(n+2\gamma_{2}+1)}\right]
\label{3.35}
\end{eqnarray}
and further, after use is made of Eq.\ (\ref{2.42}), as
\begin{eqnarray}
\sigma_{\mathrm{M},2} &=& \frac{2Z\alpha^{2}}{9}
\frac{\Gamma(\gamma_{2}+\gamma_{1}-1)\Gamma(\gamma_{2}+\gamma_{1}+2)}
{(\gamma_{2}-\gamma_{1})(\gamma_{2}-\gamma_{1}+1)
\Gamma(2\gamma_{1}+1)\Gamma(2\gamma_{2}+1)}
\nonumber \\
&& \times
\left[
{}_{2}F_{1}
\left(
\begin{array}{c}
\gamma_{2}-\gamma_{1}-1,\gamma_{2}-\gamma_{1} \\
2\gamma_{2}+1
\end{array}
;1
\right)
+(\gamma_{2}-\gamma_{1})
\,{}_{2}F_{1}
\left(
\begin{array}{c}
\gamma_{2}-\gamma_{1}-1,\gamma_{2}-\gamma_{1}+1 \\
2\gamma_{2}+1
\end{array}
;1
\right)
\right].
\nonumber \\
&&
\label{3.36}
\end{eqnarray}
According to Eq.\ (\ref{2.44}), the first of the two $_{2}F_{1}$
functions appearing above is
\begin{equation}
{}_{2}F_{1}
\left(
\begin{array}{c}
\gamma_{2}-\gamma_{1}-1,\gamma_{2}-\gamma_{1} \\
2\gamma_{2}+1
\end{array}
;1
\right)
=\frac{\Gamma(2\gamma_{1}+2)\Gamma(2\gamma_{2}+1)}
{\Gamma(\gamma_{2}+\gamma_{1}+1)\Gamma(\gamma_{2}+\gamma_{1}+2)},
\label{3.37}
\end{equation}
while the second one has been already encountered in the course of
evaluation of $\sigma_{\mathrm{E},-2}$ and is given by Eq.\
(\ref{2.45}). Hence, once Eqs.\ (\ref{2.45}) and (\ref{3.37}) are
inserted into Eq.\ (\ref{3.36}), we eventually arrive at the
following simple representation of $\sigma_{\mathrm{M},2}$:
\begin{equation}
\sigma_{\mathrm{M},2}
=\frac{2Z\alpha^{2}}{27}\frac{\gamma_{1}+2}{\gamma_{1}+1}.
\label{3.38}
\end{equation}

From the expressions for the components $\sigma_{\mathrm{M},-1}$ and
$\sigma_{\mathrm{M},2}$ of $\sigma_{\mathrm{M}}$, displayed in Eqs.\
(\ref{3.32}) and (\ref{3.38}), we deduce that the total magnetic
shielding constant for the hydrogen-like atom in its ground state is
\begin{equation}
\sigma_{\mathrm{M}}=-\frac{2Z\alpha^{2}}{27}
\frac{4\gamma_{1}^{3}+6\gamma_{1}^{2}-7\gamma_{1}-12}
{\gamma_{1}(\gamma_{1}+1)(2\gamma_{1}-1)},
\label{3.39}
\end{equation}
in agreement with the findings of Moore \cite{Moor99}, Pyper and
Zhang \cite{Pype99} and also of Ivanov \emph{et al.\/} \cite{Ivan09}
(after one takes into account that the latter authors define
$\sigma_{\mathrm{M}}$ with the opposite sign). The expression for
$\sigma_{\mathrm{M}}$ given by Zapryagaev \emph{et al.\/} in Refs.\
\cite{Zapr74,Zapr81,Zapr85} becomes equivalent to the one in Eq.\
(\ref{3.39}) after in the denominator of the former one replaces
$\lambda_{2}$ by $\lambda_{1}$, the latter quantity being identical
with our $\gamma_{1}$.

In the nonrelativistic limit one has
\begin{equation}
\sigma_{\mathrm{M},-1}\simeq\frac{2Z\alpha^{2}}{9}
\left[1+4(Z\alpha)^{2}\right]
\label{3.40}
\end{equation}
and
\begin{equation}
\sigma_{\mathrm{M},2}\simeq\frac{Z\alpha^{2}}{9}
\left[1+\frac{1}{12}(Z\alpha)^{2}\right],
\label{3.41}
\end{equation}
and also
\begin{equation}
\sigma_{\mathrm{M}}\simeq\frac{Z\alpha^{2}}{3}
\left[1+\frac{97}{36}(Z\alpha)^{2}\right].
\label{3.42}
\end{equation}
\section{Conclusions}
\label{IV}
\setcounter{equation}{0}
In this paper, we have proved that the analytical technique based on
the Sturmian expansion of the generalized Dirac--Coulomb Green
function enables one to arrive at the same simple closed-form
representations of the electric ($\sigma_{\mathrm{E}}$) and magnetic
($\sigma_{\mathrm{M}}$) dipole shielding constants for the Dirac
one-electron atom in its ground state as have been given previously
by several other authors. We find this encouraging in the perspective
of our planned, mathematically much more challenging, application of
the same calculational technique to the evaluation of
$\sigma_{\mathrm{M}}$ for an \emph{arbitrary\/} state of that atom.
Accomplishment of this task would be undoubtly worthy of effort since
although results for some particular excited states are known (cf.,
e.g., Ref.\ \cite{Moor99,Pype99,Ivan09}), in the most general case,
to the best of our knowledge, calculations of
$\sigma_{\mathrm{M}}$ have never been carried out.
%
%\newpage
%

%
\end{document}